# Rapid micro-immunohistochemistry

Robert D. Lovchik[1], David Taylor[1,2], Govind Kaigala[1*]

[1] IBM Research Europe, Saeumerstrasse 4, CH-8803 Rueschlikon, Switzerland
[2] Current address: Eidgenössische Technische Hochschule Zürich, Department of Mechanical and Process Engineering,
Sonneggstrasse 3, 8092 Zurich, Switzerland
[*]gov@zurich.ibm.com

## Abstract

We present a new and versatile implementation of rapid and localized immunohistochemical staining of tissue sections. Immunohistochemistry (IHC) allows to detect specific proteins on tissue sections and comprises a sequence of specific biochemical reactions. For the rapid implementation of IHC, we fabricated horizontally oriented microfluidic probes (MFP) with functionally designed apertures to enable square and circular footprints, which we employ to locally expose a tissue to time-optimized sequences of different biochemicals. We show that the two main incubation steps of IHC protocols can be performed on MDAMB468-1510A cell block sections in less than 30 min, comparing to incubation times of an hour or more in standard protocols. IHC analysis on the timescale of tens of minutes could potentially be applied during surgery, enabling clinicians to react in a more dynamic and efficient manner. Furthermore, this rapid IHC implementation along with the conservative usage of the tissue has strong potential for the implementation of multiplexed assays, which allows to explore optimal assay conditions on a small amount of the tissue – ensuring high-quality staining results for the remainder sample.

# 1. Introduction

Immunohistochemistry (IHC) is a well-established technique to detect the spatial distribution of antigens in tissue sections and is now routinely used for tumor analysis in research laboratories and diagnostic centers. Interestingly, this method has been used for well over 50 years with limited modifications to the underlying principle itself, owing to its robustness. In contrast, it has undergone several innovations regarding implementation, automation and throughput. With each innovation, user input has been reduced, but often at the expense of flexibility and versatility.

After sample-dependent pre-treatment, in the first step of an IHC assay a primary antibody, which specifically binds to an antigen of interest, is presented to the tissue. In the second step, a secondary antibody is applied. The secondary antibody, which helps to amplify the detection signal, is usually polyclonal and conjugated with enzymes that enable a subsequent detection of the signal through colorimetric reactions. This series of incubation steps allows for antigen localization through choice of antibody and control over the reaction kinetics, thus aiding semi-quantitative or quantitative estimation of the amounts of antigen present in the sample. The commercially sourced antibodies in these protocols often vary in their specificity and sensitivity, thus requiring meticulous optimization and testing[1]. Hence, there is a need to enable users to rapidly screen parameter spaces, in order to determine practical assay conditions for the specifically used bio-chemicals. The ability to *quantitatively characterize* detected signals, perform *multiplexed detection of various antigens simultaneously*, and to *rapidly* implement IHC protocols that are standardized or modified as per user need, are facets of research that would be essential to fostering the next generation of methods in cancer research and diagnostics.

Several notable efforts have been made to enable a quantitative proteomic assessment of tissue sections via IHC, for example, Dupouy *et al.* [2], Carvajal-Hausdorf *et al.* [3], and Xing *et al.* [5]. In our previous work, Kashyap *et al.* [4], using a microfluidic probe, we introduced a method towards the implementation of IHC as a quantitative assay (qµIC). Our qµIC method looks at the evolution of the signal strength to extract quantitative information. The current design of the MFP devices that we present here could also be used for qµIC. To broaden the applicability of the MFP, this work shows how we increased the rapidity of IHC and implemented sequential chemistry in small and defined regions of tissues. Significant improvements pertaining to the speed and reproducibility of these assays remain essential for rendering quantitative IHC useful in a routine clinical context.

Numerous methods have also been proposed for implementing multiplexed IHC analysis. These are broadly classified as: 1. *Application of different primary antibodies to the same region of a tissue*. To avoid cross-reactivity in the detection step, primary antibodies must be produced in different species, which strongly limits the number of antigens that can be detected simultaneously [5]. This constraint can be relaxed by stripping antibodies deposited in a previous analysis and by bleaching the resulting detection signal prior to performing a subsequent test. This sequential stripping and bleaching is laborious, time-consuming and, importantly, tends to deteriorate the tissue. 2. *The layered peptide arrays technique*, where the antigens are blotted onto several layers of stacked membranes, allowing for independent analysis of each membrane subsequently, while preserving the spatial information of the obtained IHC data [6]. Although suitable for multiplexing, this approach is complex and consumes the tissue section entirely. 3. *Spatial segregation to present different primary antibodies on a tissue section*, allowing to use any type of antibodies on neighboring reaction sites [7]. Depending on the resolution of the spatial confinement of the corresponding antibody solutions, this allows the mapping of specific regions of interest (ROIs) in a multiplexed manner. This implementation overcomes several of the limitations of other implementations described above, by obviating harsh chemical stripping or adding complexity to the IHC protocol.

One potential strategy towards spatially segregated multiplexing is to implement co-flow laminar streams of antibody solutions in open-top microchannels pressed onto a tissue section, as reported by Kim *et al.*[8] and Ciftlik *et al.*[9]. The required mechanical contact with the tissue sections nevertheless consumes parts of the sample and offers limited flexibility in terms of addressing specific ROIs on the tissue section, which becomes critical when ROIs are very small in comparison with the total tissue area. A versatile and IHC compatible strategy is to localize antibody solutions in a so-called hydrodynamic flow confinement (HFC) formed at the apex of a non-contact liquid scanning probe, such as the microfluidic probe (MFP)[10]. Multiplexed IHC routines using the MFP were demonstrated by Lovchik *et al.* [7] and Queval *et al.* [11] at the µm-length scale and expanded to the cm-scale by Taylor *et al.* [12]. Such flow-based approaches were also used for brain slice microenvironment modification [11], breast cancer sections [12], and in combination with lanthanide-based immunocomplexes [13] and immunofluorescence [14].

Another important aspect pertaining to IHC analysis is the total time required to implement the complete assay. In traditional IHC, the incubation time of each antibody is approximately 30-60 min, sometimes longer, even overnight. This step significantly contributes to the total time required for an IHC test. In earlier work, we presented micrometer-scale IHC[7], in which we implemented highly localized delivery of primary antibodies on tissue sections with an MFP. The time required for the presentation of the primary antibody to each area of the section was in the range of about 30-60 s. Such a microfluidic approach, allowing a reduction of the incubation times, was enabled by two advantages: (a) only small volumes of liquids are consumed, allowing to increase the concentration of antibodies without increasing the cost of the assay, and, (b) the flow of the antibody solution allows for continuous replenishment of the antibodies consumed from the solution, thereby making the reaction time largely independent of the relatively low diffusion coefficients of the antibodies. Apart from our demonstration of expanding MFP assisted IHC from the µm-scale to the cm-scale, Taylor *et al.* [12], a related publication by Cors *et al.*[15] additionally showed the integration of dewaxing and rehydration with an MFP to further increase the rapidity of multiplexed IHC staining.

The challenge faced in our and others' prior work was that processing the whole slide for the standard IHC workflow, with the MFP performing only one of the steps, made later use of the slides for other assays challenging. In this paper, we focus on taking advantage of the microfluidic probe to boost the rapidity of the assay by enhancing the reaction rates and automation by rapid switching between different types of processing liquids, allowing for the fast implementation of sequential assay steps (see Fig.1.). We further demonstrate different ways of performing sequential chemistry with an MFP to integrate the whole workflow, such that other assays might be run concomitantly, prior to any whole slide processing. The rapidity of the assay and localization of the applied liquids allow to optimize the processing parameters efficiently, using only a small fraction of the tissue. We designed a new horizontally oriented MFP head with apertures that enable the implementation of arbitrarily shaped footprints (see Fig. 2.). Radial and square shaped flow confinements were chosen, which are well suited for sequential presentation of different liquids on surfaces from the µm to mm-scale. Such radial or square shaped HFCs are formed by central injection of a processing liquid, which is re-aspirated by surrounding ring or square-shaped apertures. We chose to locally implement the three core steps of an IHC routine using horizontal probe heads: the application of the primary antibody, the secondary antibody and the enzyme (peroxidase). These three steps determine to a large extent the quality of the obtained stain and usually require 30 min up to several hours in on-bench protocols. Due to convection-enhanced deposition and the possibility to use higher concentrations of antibodies without increasing the overall consumption of reagents, the MFP is well suited to locally implement these core steps of IHC analysis within minutes - we refer to this method as "rapid µIHC".

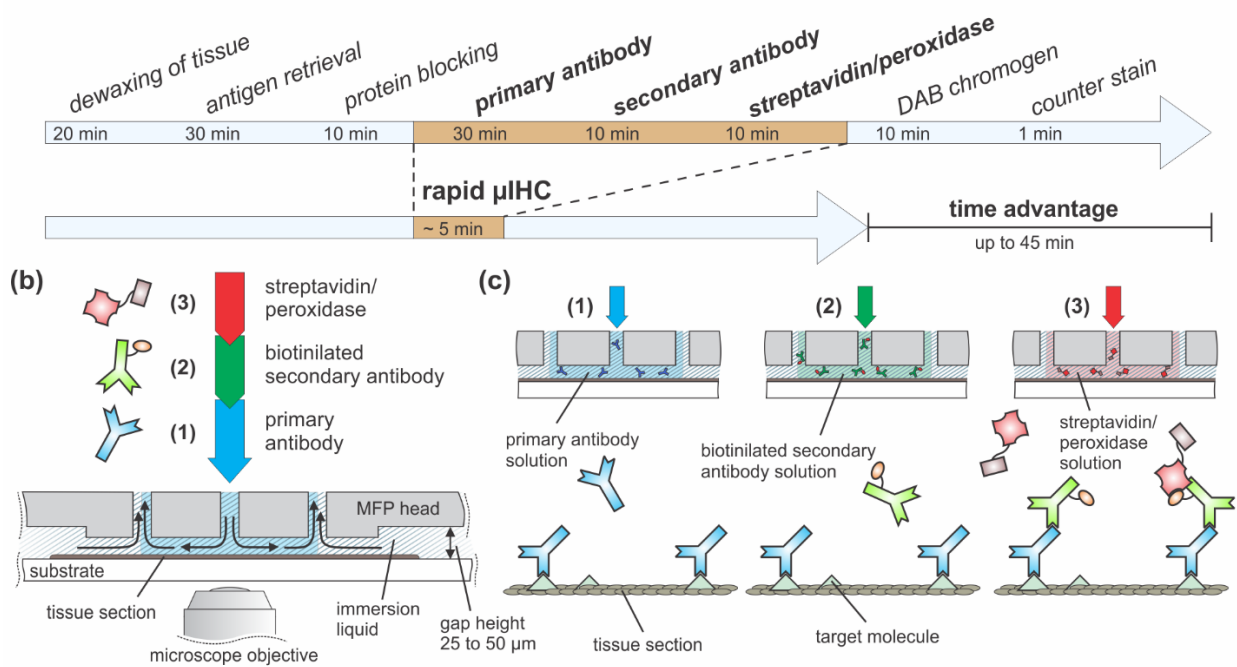

*Figure 1. a) Protocol for immunohistochemistry implemented with chromogen-based revelation. Rapid µIHC is enabled by implementing the three core steps of the IHC protocol using an MFP. The times shown for the individual steps of the conventional workflow are taken from the suggested protocol of the Mouse specific HRP/DAB detection kit from Abcam (ab64259). b) Scheme of three different reagents passing through the central aperture of an MFP head placed over a substrate.*

*The central injection aperture is linked to three channels through which the reagents are injected sequentially. c) The three core steps of rapid µIHC are (1) the binding of a primary antibody to an antigen on the sample, (2) binding of a biotinylated secondary antibody to the primary antibody, and, (3) applying a streptavidin/peroxidase solution for revelation.*

## 2. Theory

Important steps of immunohistochemistry staining protocols are the incubation of the tissue with primary and secondary antibodies. Here, we show the dependency of the IHC signal on multiple parameters. The amount of antibodies binding to receptors on the surface of the sample $b(t)$ depends on several parameters. The value of interest, to be semi-quantitatively assessed by an IHC staining routine, is the surface density of binding sites $b_m$. Further factors that can be adjusted to optimize the assay conditions of an MFP-based staining protocol are the composition of the buffer, the temperature $T$, the bulk concentration of the antibodies in the processing liquid $c_0$, the flow velocity of the processing liquid (a function of the injection and aspiration flow rates, $Q_i$ and $Q_a$, and the gap distance $d$) as well as the incubation time $t_{inc}$. In contrast, the intrinsic properties of the applied antibodies, namely the association and dissociation constants with their respective target $k_{on}$, $k_{off}$ and their diffusivity $D_{ab}$, are not directly accessible to optimization and are subjected to significant batch-to-batch variations and variability induced by varying storage conditions and other factors [16][17].

To estimate the impact of a change in $k_{on}$, e.g. of the primary antibody, we assume that, due to convection inherent to the MFP, the binding of antibodies to the sample is reaction limited and $D_{ab}$ can therefore be neglected. We also assume that all the involved species are first order reactants and the receptor ligand binding reaction at the sample therefore follows a Hill-Langmuir characteristic. In such a scenario, the rate of the surface reaction is

$$\frac{db}{dt} = k_{on} \cdot c \cdot (b_{tot} - b) - k_{off} \cdot b \qquad (1)$$

Integration of (1) yields the amount of primary antibodies bound to the sample after a given time $b(t)$

$$b(t) = b_m \cdot \frac{c}{k_{off}/k_{on} + c} \cdot \left(1 - e^{-(k_{on} \cdot c + k_{off}) \cdot t}\right) \qquad (2)$$

The impact of a reduction of $k_{on}$ is demonstrated by estimating how much additional time $t_{add}$ would be needed to reach saturation in a reaction $b_{slow}(t)$ with a lower $k_{on}$ compared to a reaction $b_{ref}(t)$ with a constant $k_{on}$. Assuming $k_{on,ref} = 1 \cdot 10^6 \text{ M}^{-1}\text{s}^{-1}$ for the reference reaction and $k_{on,slow} = 0.5 \cdot 10^6 \text{ M}^{-1}\text{s}^{-1}$ and an antibody concentration of 25 µL/mL for both reactions, we find that in the reaction with a lower $k_{on}$ saturation of the surface (95%) is delayed by about 110% in comparison to the reference case.

As the variation in the performance of applied antibodies is found to have a large impact on the outcome of IHC assays, there is a strong need for a reliable way to validate antibodies under representative conditions to ensure consistency of results [17].

# 3. Materials and Methods

## 3.1 Microfluidic probe head fabrication

The microfluidic heads are hybrid Si/glass devices fabricated using a sequence of standard microfabrication steps: photolithography, deep reactive ion etching (DRIE), anodic bonding and dicing. The design of the MFP heads was created in L-Edit V2016.2 (Tanner EDA) and the masks for photolithography written on a laser writer (DWL 2000, Heidelberg Instruments, Germany). DRIE (AMS-200SE, Alcatel Micro Machining Systems, France) was performed on a double side polished 4-inch Si wafer (500 µm thick, Siltronix, Geneva Switzerland) to pattern the microchannels, the through-etched apertures and the mesa of the MFP heads. The microchannels were etched to a depth of 50 µm. All three etching steps also contributed to patterning the outline of the hexagonal shape of the MFP heads. Si bridges were left to stand around the outline to keep the parts in place for further process steps (more details in the SI). A 4-inch glass wafer (1 mm, Borofloat 33, SCHOTT AG, Germany) was used to fabricate the upper part of the head. Prior to anodically bonding the glass wafer to the patterned Si wafer, 500 µm vias were drilled on a CNC milling machine (Wissner Gamma 303, Goettingen Germany) using a diamond drill (Haefeli AG, Switzerland). Wafer-level anodic bonding took place on a custom-made bonding machine (475 °C, 1.3 kV). After dicing, the temporary Si bridges were removed using a scalpel blade.

## 3.2 Fabrication and setting up the microfluidic probe station

A custom holder was built to mount the MFP head and connect it to the pumping system, Fig 2. The holder is assembled from several machined pieces and is attached to the Z-axis of a XYZ scanning system, which is placed on an inverted microscope (Eclipse Ti, Nikon Instruments, Japan), see SI1. Mounting the MFP head was done in the following way: (1) the head was positioned in the holder (aluminum) and (2) the PDMS gasket (molded from Sylgard® 184) placed on the head. The PDMS gasket self-aligned with the holder to match the holes to the via in the head. (3) A pressing block (PMMA), comprising holes for the capillaries, was placed on top of the gasket. The capillaries were inserted at the positions matching the microchannels in the head. (4) A bridge comprising a resilient pressure piece was mounted to the frame of the holder and the resilient pressure piece was screwed toward the pressing block (finger-tight). This force compressed the gasket to lock the capillaries in place and ensure a leak-free connection, Fig 2 and SI3.

With the MFP head mounted, the system was aligned to ensure co-parallelism between the scanning stage XY movement, the substrate and the head surface. This was done by placing a dummy glass slide (Menzel, Germany) and adjusting the tilt of the head and substrate while establishing the zero Z-position at different positions on the slide repeatedly. Finding the zero Z-position was done by slowly moving the head towards the glass slide and observing the appearance of Newton's rings between the head's apex and the glass.

### 3.3 Finite-Element Modeling

We performed steady-state 3D simulations with COMSOL Multiphysics (version 4.2). Nonslip boundary conditions were defined on all surfaces, constant flow velocity boundary conditions were applied at the aperture and a constant pressure boundary condition was applied at the virtual interface between the liquid in the gap and the surrounding liquid at the edges of the apex. All fluids were set to be water (incompressible Newtonian fluid with a density of 998 kg/m$^3$ and a dynamic viscosity of 0.001 N s/m$^2$). The ratio of aspiration flow to injection flow was kept at 2.5, with an aspiration flow rate of 10 µl/min.

### 3.4 Micro-immunohistochemistry protocol

(1) Sample preparation: MDAMB468-1510A cell block sections were purchased from AMS Biotechnology Europe (Massagno, Switzerland). A standard protocol was applied to the cell block sections to prepare the sample for the IHC experiment. The preparation involved deparaffinization, hydrogen peroxidase block, target retrieval and protein block (see SI4 for more details on the protocol). The prepared cell block sections were kept in PBS at 4 °C no longer than one day until use.

(2) MFP system setup: Two glass slides were placed into the substrate carrier next to each other. This allowed adjusting the fluidic system to achieve stable flow confinements on a dummy glass slide, before performing experiments on the sample glass slide. The injection channels of the system were initially purged with colored liquids (diluted food colorant) until stable flow confinements were observed on the dummy glass slide. The pressure settings for each channel were saved in the control software to enable rapid switching between flow confinements with different reagents.

(3) µIHC: The MFP head was moved to the sample glass slide, which was covered with PBS, and positioned away from the cell block section at a safe distance (e.g. 15 mm). The substrate to head gap distance was set to 75 µm. The reservoir tubes containing food colorant were replaced with tubes containing the reagents for IHC (primary antibody, secondary antibody and for some experiments streptavidin peroxidase and DAB substrate). The primary antibody (ab122898, anti P53) as well as the other IHC chemicals (Mouse specific HRP/DAB (ABC) Detection IHC Kit, ab64259) were purchased from Abcam (Cambridge, UK). Each injection channel was purged with the according IHC reagent until no residual food coloring could be observed. The pressure settings for stable flow confinements and switching between liquids were adjusted if needed. The MFP head was then moved to the X and Y zero position in relation to the cell block section (upper left corner of an imaginary square fitted around the circular cell block section). This starting point for the experiment was set to 0 for the X and Y axis in the control software.

(4) Design of experiments for micro-IHC conditions: Rapid µIHC with varying incubation times of the reagents was performed by moving to designated coordinates over the cell block section and switching between the reagent injection channels while keeping the aspiration flow constant. For scanning µIHC, the MFP head was moved along the row of flow confinements at constant velocities. After applying the reagents with the MFP, the sample was processed for the subsequent steps according to a conventional IHC method. Depending on the number of reagents applied with the MFP, the cell block sections were incubated with streptavidin peroxidase and DAB substrate or only DAB substrate. No further treatment was needed for the cell block sections after scanning µIHC. Processed cell block sections were dehydrated and mounted using standard mounting medium and a coverslip.

### 3.4 Imaging and data processing

An upright microscope (Eclipse 90i, Nikon Instruments, Japan) was used to image the sections and color photographs were taken of each sample with 4× and 10× magnification. Quantification of the staining intensity was done using ImageJ. Features were extracted manually as this was considered sufficient for the conceptualization shown here. The individual images were analyzed as follows: (1) at least 25 nuclei per spot were selected with the freehand selection tool and the mean gray value calculated from all areas, (2) the same procedure was performed to calculate the mean gray value of the cytoplasm of the cells, where at least 25 regions per spot were selected manually for analysis. The normalized stain intensity per spot was then determined through subtraction of the two values.

## 4. Results and discussion

### 4.1 Horizontally oriented microfluidic probe heads for sequential chemistry

In Lovchik *et al.*[7], we demonstrated multiplexed µIHC on tissue sections within a single core of a tissue microarray. In that work, we implemented a single step of the µIHC process by locally presenting the primary Ab solution to the tissue with a vertically oriented MFP head, using the basic configuration of an HFC formed via two square-shaped apertures. While this was a stride forward in performing local µIHC, a significant challenge remained of how to sequentially present multiple liquids to a small region of a tissue to implement a complete IHC protocol and how to increase the rapidity in the implementation of the complete protocol. The ability to switch between reagents can of course be extended to alternating between different primary antibodies for multiplexed analysis. In contrast, in this paper, we propose a new class of horizontally oriented MFP heads that allows not only to elegantly implement efficient sequential protocols using multiple processing liquids, but also provides flexibility in the placement of the apertures relative to each other and offers freedom in designing the apertures. The horizontal arrangement of the new generation of MFP heads provides room for the fluidic routing required for switching between different processing liquids and allows to deliberately shape apertures to enable the creation of different shapes of the confined liquids.

A challenge related to the horizontal orientation of the MFP heads is the resulting large surface area of the probe head. It increases the risk of damaging the sample due to improper alignment of the probe head with reference to the tissue section. We mitigate this issue by forming a 50 µm high mesa at the center of the probe head, in which the apertures are placed, and which is then closer to the sample, while the distance to the sample is larger for the remaining area of the probe head. There are several noteworthy aspects to the new head designs: (*i*) different geometries of the flow confinement are feasible on account of the flexibility of placement and shape of the apertures (square, circle etc.). (*ii*) The circular arrangement of the apertures helps to minimize the cross-influence between adjacent HFCs, for which the earlier approach was to have additional stabilization apertures/flows, Taylor *et al.*[12].(*iii*) Due to central injection and side aspiration, the device can work in a more stable manner for larger substrate-to-head distances.

The spacing between the apertures was designed to minimize cross-talk between the different flow confinements when running simultaneously. This was done by performing finite element simulations for both the circular and square-shaped HFCs (see Fig. 2c). An important aspect of the simulations was to ensure that the HFCs are in contact throughout the entire desired processing footprint on the sample. The simulations clearly showed that this is the case for gap distances between 50 and 100 µm, which is the range used in this work.

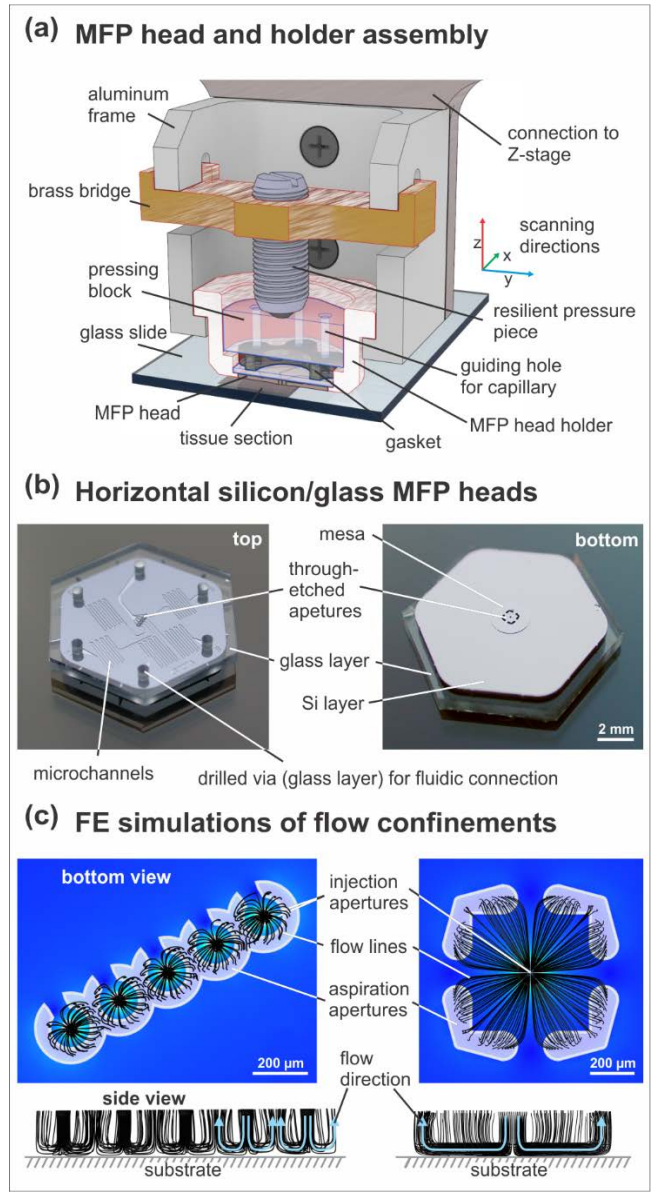

*Figure 2.* The holder assembly, single units of horizontal microfluidic probe heads, and (corresponding) FE simulations of flow confinements. a) Assembly of the MFP head holder: The MFP head is seated on its glass overhang within an aluminum holder. A molded PDMS gasket for holding and sealing up to 6 capillaries is placed on top of the head and a mechanical compression force is applied to the gasket using a resilient pressure piece and a pressure block. All components of the assembly are held by an aluminum frame which is mounted onto the Z-stage of the scanning unit. b) Photographs of horizontal MFP heads made of Si and glass. The top view shows the vias drilled through the glass layer for fluidic connection and the routing of the microchannels for addressing the apertures of the head. The mesa and the injection/aspiration apertures can be seen on the bottom view. The height of the mesa is typically 50 µm and the diameter varies between 2 and 4 mm, depending on the aperture designs. c) FE simulations were used to study injection/aspiration ratios for different HFC geometries at varying heights.

## 4.2 Implementing sequential chemistry using the MFP

We propose and implement two approaches to sequentially expose a surface to different reagents in order to implement a complete biochemical assay. One approach is to sequentially deliver liquids with the MFP positioned in a specific location over the surface. The second one is to perform the sequential reagent exposure of the liquids to the surface by leveraging the scanning ability of the MFP in combination with several neighboring HFCs.

In the liquid switching sequential chemistry approach, the MFP head comprises three connectors to different liquids that merge into one injection aperture that opens into the center of the MFP head. By splitting the aspiration into four L-shaped structures as seen on the bottom face of the head (see Fig. 3a), a square-shaped flow confinement is achieved. Here we demonstrate sequential chemistry in a defined region by switching the flow of reagents externally. For this, we control the flow of each individual injection channel. Since the merging of the three channels is close to the injection aperture, the dead volumes are extremely small (<1 µL) and therefore the switching times for typical injection flow rates (~6 µL/min) is just a few seconds. This number takes into account the exchange of the total liquid volume within the flow confinement (0.02 µL). Because the subsequent reagent completely displaces the previous liquid, intermediate washing steps, as applied in conventional protocols, are eliminated. This adds to the time advantage of this implementation. Using the approach of exchanging the liquids in the HFC, we performed anti-P53 staining on a 4 mm diameter cell block section (MDA-MB468). We implemented three steps of the IHC protocol, wherein primary antibodies, secondary antibodies and streptavidin peroxidase are displayed to the cell block section sequentially. As seen in Fig. 3b, we stained footprints of roughly $500 \times 500$ µm$^2$ and in each of these footprints we varied the incubation times of the three steps (incubation times provided in fig. SI.5). As expected, the signal intensities differed depending on the incubation strategies and good contrast and uniformity was obtained even with incubation times of the primary antibody as low as 30 s.

Another approach to perform rapid sequential exposure of multiple liquids on a surface is to scan with an MFP multiple HFCs along one axis over the surface. We designed a specific MFP head, where 5 HFCs can be run simultaneously in a linear arrangement, each one confining a different liquid. Fig.3 c and d show the experiment performance using four different chemicals, with the fifth port left unused. The size of the HFCs and the scanning velocity define the incubation times for the reagents on the surface of the sample. Note that the incubation time of an individual reagent could also be adapted through the chip design or by confining the same reagent in multiple neighboring HFCs. The injection channels in this configuration do not merge in a single injection aperture as in the other chip design. This is an advantage, because two sequential reagents could potentially react and clog the channel. As each flow confinement is 500 µm wide, and these confinements are linearly arranged during scanning, the resulting stained tissue section appears as a line along the y-axis, Fig. 3d. In addition, the separation of the HFCs prevents cross contamination exchange of reagents, because there is a small amount of immersion liquid between all HFCs as illustrated in the bottom of Fig. 2c. The scanning approach also led to good staining results with an incubation time of 30 s for each of the four implemented steps.

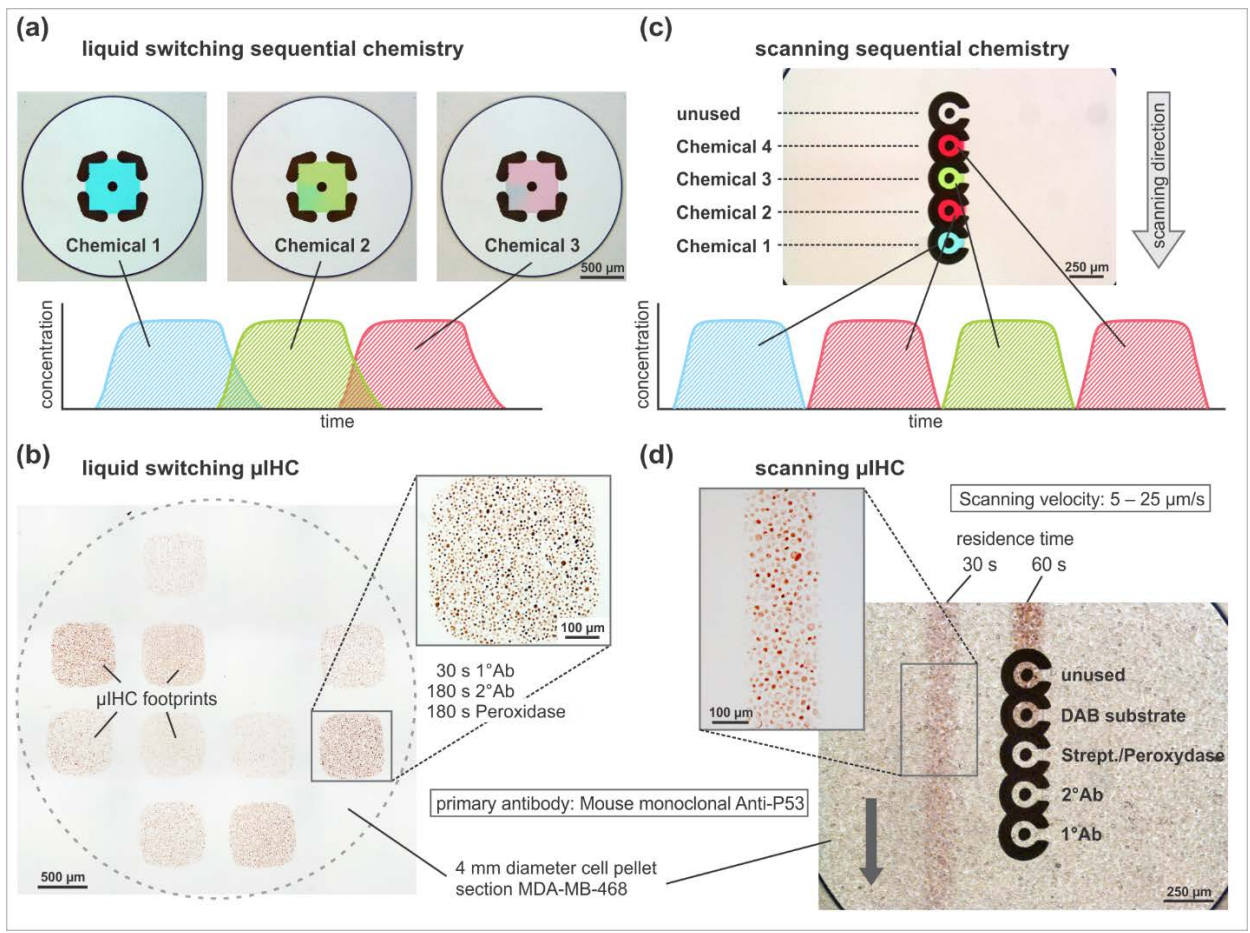

*Figure 3.* Experimental results of two implementations of sequential chemistry to perform the complete IHC protocol on cell blocks. (a) and (b) are with the head static, while (c) and (d) are with the head scanning. a) Photographs of a single HFC captured at three different times with different colored liquids. The liquids within the HFC can be switched within about 2 seconds. b) IHC staining for P53 on a cell pellet section with varying incubation times of the primary Ab, secondary Ab and streptavidin peroxidase at different locations. c) Scanning of the reagents with a multi-HFC head allows fast sequential chemical treatment of a surface. This strategy was also applied for IHC on a cell pellet section (d) and good contrast was achieved with residence times as short as 30 s.

## 4.3 Towards optimizing processing parameters for establishing rapid and efficient IHC

In conventional IHC the practice is to make use of protocols which previously provided good results for a given type of tissue, antigen of interest and primary antibody. Predefined protocols, however, do not take into account the expression level of the antigens in a specific sample, or batch-to-batch variations in the performance of the applied antibodies. A simple and rapid way to first test the IHC parameters for a given sample would therefore allow to optimize the quantity and the quality of the information that can be retrieved from an IHC analysis. It is possible to perform such optimization using entire tissue sections. However, in order to traverse a reasonable number of process parameters, the amount of tissue required to determine an ideal set of processing parameters will be rather large. As the MFP allows to perform IHC at the micrometer length scale, a fraction of the section can be dedicated to such optimization purposes. Alternatively, instead of using fractions of several sections, even a single section can be dedicated to establishing the suitable IHC process parameters. While this idea is well suited for diagnostics, it is even more suitable for optimizing conditions for screening-like applications, as are performed with tissue microarrays.

To demonstrate this concept of process parameter optimization, we made use of another cell block section for demonstrating sequential IHC. To determine the ideal incubation times for the primary and secondary antibody solutions, we performed a design of experiments (central composite design), in which these two factors were systematically varied (see Fig. SI.6). The design of experiments consisted of independent sequential reactions performed on 12 spots of the same cell block section. To ensure that there is no non-specific binding of the antibodies, we implemented negative controls without primary or secondary antibodies on two additional spots (bottom left and right spot in Fig. 4a). A selection of higher resolution images can be found in Fig. SI.7.

After spot-wise sequential incubation, the streptavidin-peroxidase and the DAB chromogen were applied over the entire section. Even with highly variable incubation times stained areas may appear very similar at first glance. However, it is essential to determine the difference in the staining intensities between the nuclei and the cytoplasm to extract meaningful quantitative data. An analysis of the results revealed significant first and second order effects of the incubation time with primary and secondary antibodies (Fig. 4b). While increasing either of the two incubation times had a positive effect on the signal level, the signal contrast deteriorated for longer incubation with primary antibodies. Our results therefore suggest that for the current set of antibodies/antigens an optimized workflow should have a short incubation time for the primary antibody and an incubation time about five times longer for the secondary antibody. While the general idea of having minimal incorrect binding in the first step (primary antibody) and high amplification in the second step holds, the exact ratio of incubation times during each step might vary and would need to be established on a case by case basis.

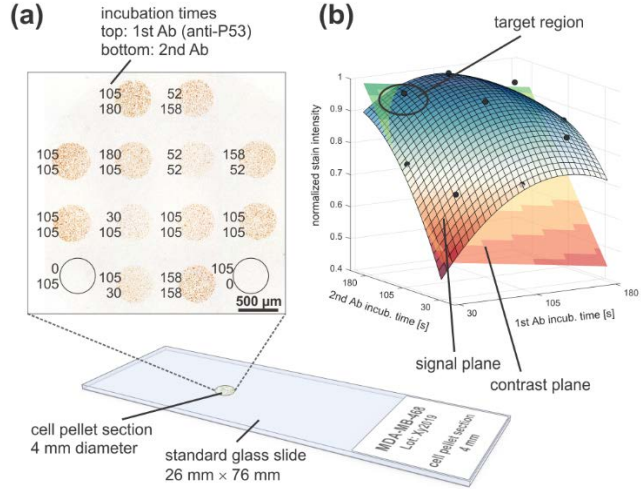

*Figure 4. Exploring suitable staining conditions using process parameter optimization. a) Experimental result of a two-factor central composite design with two negative controls on a single cell block section (incubation times shown in seconds). Streptavidin-peroxidase and DAB chromogen were applied globally. b) Analysis reveals statistically relevant first and second order effects for both factors. The figure displays the response surfaces for the signal intensity (red-white-blue) and for the signal contrast (red-yellow-blue). The negative curvature of the intensity response surface for long incubation times with primary antibodies does not make sense physically and is considered an artifact, which could be removed by increasing the number of datapoints. The signal contrast deteriorates for long incubation with the primary antibody. Therefore, relatively short incubation with primary antibodies and relatively long incubation with secondary antibodies results in the highest signal at best contrast.*

## 5. Summary and concluding remarks

The presented strategy for rapid IHC analysis is conservative in the consumption of a tissue section, benefits from reduced amounts of reagents and is able to screen the process parameters of the IHC protocol for good quality stains. This approach complements our previous work on computational heuristics to establish process parameters for high-quality staining [18][19]. At the core of performing such micro-scale IHC reactions on a tissue section is hydrodynamic confinement of liquids using the microfluidic probe. Here, we specifically propose two designs of MFP heads to implement a complete IHC protocol. The implementation of the µIHC protocol is rapid in the first case through the continuous replenishment of the reagents on the tissue surfaces due to the flows inherent to the MFP [20] and in the second case, due to the sequential presentation of reagents, which also simultaneously rinse out the previous liquid. From a technical perspective, this new class of MFP heads provides new opportunities of how liquids can be presented to the surface and how explicit shapes of the confined liquid can be formed.

Due to the above attributes, we think this may be the genesis of real-time processing of tissue sections with an immediate readout using non-fluorescent labels, thus being compatible with existing pathology workflows. The devices and methods presented build on established and robust workflows, while significantly enhancing their dynamics and versatility. Also, real-time IHC staining can be readily applied to create gradients of the primary antibody and through this quantitative IHC becomes possible [21]. The strategies presented here can be used for whole tissue processing. Furthermore, to capture the variations in the tissue, and support statistics-guided decision-making, we think an image processing software would facilitate the implementation of MFP-based IHC into fast and automated analysis routines. The presented methods may be extended to also include pre-analytical steps required prior to staining (such as e.g. de-waxing and antigen retrieval treatments).

Another interesting direction of this approach is to study the kinetics of an antibody-antigen reaction. It is well known that applying mathematical models does not allow to accurately establish the association constants of the applied antibodies. Nonetheless, our method for the local implementation of a specific combination of staining conditions reveals the impact of varying $k_{on}$ and compares the performance of different batches to optimize protocols in a targeted manner. The incubation times of primary and secondary antibodies can be varied systematically, following a design of experiments and creating arrays of stains representing different incubation conditions. In this work, we implemented a full factorial design to assess the effects of the incubation times with the primary antibody and the secondary antibody. Such an array of stains featuring permutations of different incubations times with the primary and secondary antibody solutions would allow to compare antibodies from different batches or vendors and to optimize incubation times in staining protocols in a targeted manner. In summary, our technology can be used for quality checks and screening, and we think that it has the potential to provide deeper insight into precious tissues and more information within scientific studies. Moreover, we see large potential in implementing rapid IHC in scenarios where time to result is highly critical.

The current development stage of the MFP heads and the experimental platform still has some limitations. The stability of the liquid flow and switching between liquids is sensitive to air bubbles forming in the tubing or channels. Although degassing of liquids and good practice in setting up the experiments minimizes the occurrence of bubbles and consequent failure of the experiment, there is still need for some further engineering of the system. There are several strategies in microfluidics to avoid bubble formation as well as coping with them [22]. We are also aware that the current design is complex for fabrication, especially the Si/glass hybrid heads that require microfabrication in a clean room. We are currently translating several of the MFP head designs to work in plastics and become suitable for injection molding.

In summary, we think in pathology, there is a need for strategies to enable pre-operative molecular screening of tissues while using as much of the current workflows as possible. IHC, due to its complex multi-step and sometimes even multi-day protocol, is confined to central laboratories. Increasing its rapidity and automation could specifically help this method to be used on diagnostics on fresh tissue in combination with surgery. We believe that, although still rudimentary, the methodology presented here provides a starting point towards closing that gap, all the while enabling high quality quantitative IHC to reduce errors in diagnostics. To optimize our method for use in a clinical setting, further validation through pathologists would be required. Such input would help to determine, for example, the value of the time advantage for multiplexed tissue analyses and whether the size of the stained regions is sufficient for diagnostic interpretation. Overall, the attributes of the methodology could also help significantly in addressing open research questions like probing tumor heterogeneity.

## Supporting Information

SI.1 Fabrication of microfluidic probe heads

SI.2 Microfluidic probe platform

SI.3 MFP head holder and associated components

SI.4 Dewaxing and target retrieval protocol

SI.5 Incubation times of spots in figure 3b

SI.6 ANOVA table for design of experiments displayed in figure 4

SI.7 Photographs of anti-P53 stainings on cell block sections

Videos S1

## Acknowledgements

We thank Anna Fomitcheva Khartchenko, Aditya Kashyap and Lena Voith von Voithenberg for providing support during the experiments and Linda Rudin for critical comments on the manuscript. We thank Ute Drechsler and Diana Davila Pineda for help with the fabrication of the probe heads. We acknowledge Emmanuel Delamarche and Heike Riel for their continuous support. This work was supported by the European Research Council PoC Grant CellProbe (842790).